\documentclass[twocolumn]{aastex63}

\usepackage{graphicx}	
\usepackage{color}
\usepackage{amsmath}	
\usepackage{amssymb}	
\usepackage{mathrsfs}
\usepackage{mathrsfs}
\usepackage{multirow}

\newcommand{\msun}{M_\odot}

\newcommand{\zsun}{Z_\odot}

\newcommand{\cc}{{\rm cm}^{-3}}

\newcommand{\pc}{{\rm pc}}
\newcommand{\mum}{\mu {\rm m}}

\newcommand{\K}{{\rm K}}
\newcommand{\beq}{\begin{equation}}
\newcommand{\eeq}{\end{equation}}

\shorttitle{}
\shortauthors{Li et al.}

\begin{document}

\title{
Little Red Dots: Rapidly Growing Black Holes Reddened by Extended Dusty Flows
}

\correspondingauthor{Zhengrong Li, Kohei Inayoshi}
\email{lizhengrong@pku.edu.cn, inayoshi@pku.edu.cn}

\author[0000-0002-8502-7573]{Zhengrong Li}
\affiliation{Kavli Institute for Astronomy and Astrophysics, Peking University, Beijing 100871, China}
\affiliation{Department of Astronomy, School of Physics, Peking University, Beijing 100871, China}

\author[0000-0001-9840-4959]{Kohei Inayoshi}
\affiliation{Kavli Institute for Astronomy and Astrophysics, Peking University, Beijing 100871, China}

\author[0009-0005-1831-3042]{Kejian Chen}
\affiliation{Kavli Institute for Astronomy and Astrophysics, Peking University, Beijing 100871, China}
\affiliation{Department of Astronomy, School of Physics, Peking University, Beijing 100871, China}

\author[0000-0002-4377-903X]{Kohei Ichikawa}
\affiliation{Global Center for Science and Engineering, Faculty of Science and Engineering, Waseda University, 3-4-1, Okubo, Shinjuku, Tokyo 169-8555, Japan}
\affiliation{
Department of Physics, School of Advanced Science and Engineering, Faculty of Science and Engineering, Waseda University, 3-4-1,
Okubo, Shinjuku, Tokyo 169-8555, Japan}

\author[0000-0001-6947-5846]{Luis C. Ho}
\affiliation{Kavli Institute for Astronomy and Astrophysics, Peking University, Beijing 100871, China}
\affiliation{Department of Astronomy, School of Physics, Peking University, Beijing 100871, China}

\begin{abstract}
The James Webb Space Telescope (JWST) observations have revolutionized extragalactic research, particularly with the discovery of little red dots (LRD), which have been discovered as a population of dust-reddened broad-line active galactic nuclei (AGNs).
Their unique v-shape spectral feature characterized by a red optical continuum and a UV excess in the rest frame 
challenges us to discern the relative contributions of the galaxy and AGN.
We study a spectral energy distribution (SED) model for LRDs from rest-frame UV to infrared bands.
We hypothesize that the incident radiation from an AGN, characterized by a typical SED, is embedded in an extended dusty medium
with an extinction law similar to those seen in dense regions such as Orion Nebula or certain AGN environments.
The UV-optical spectrum is described by dust-attenuated AGN emission, featuring a red optical continuum at $\lambda>4000~{\rm \AA}$
and a flat UV spectral shape established through a gray extinction curve at $\lambda<3000~{\rm \AA}$, due to the absence of small-size grains.
There is no need for additional stellar emission or AGN scattered light.
In the infrared, the SED is shaped by an extended dust and gas distribution ($\gamma<1$; $\rho\propto r^{-\gamma}$)
with characteristic gas densities of $\simeq 10-10^3~\cc$,
which allows relatively cool dust temperatures to dominate 
the radiation.
As a result, these dust structures shift the emission energy peak from near- to mid-infrared bands in the rest frame; for sources at $z\sim 4-7$, the corresponding wavelengths shift from the JWST/MIRI to Herschel range.
This model, unlike the typical AGN hot torus models, can produce an infrared SED flattening that is consistent with LRD observations through JWST MIRI.
Such a density structure can arise from the coexistence of inflows and outflows during the early assembly of galactic nuclei.
This might be the reason why LRDs emerge preferentially in the high-redshift universe younger than one billion years.
\end{abstract}

\section{introduction}
The James Webb Space Telescope (JWST) has dramatically advanced the study of high-redshift ($z>4$) active galactic nuclei (AGNs) 
by unveiling low-luminosity AGN populations more representative than luminous quasars \citep[e.g.,][]{Onoue_2023}. 
Among these discoveries, very red and compact objects, so-called little red dots (LRDs), have attracted considerable attention 
\citep[e.g.,][]{Barro_2023,Kocevski_2023,Labbe_2023,Matthee_2024}.  
Spectroscopic observations on photometrically identified LRDs have revealed that more than 70\% exhibit broad hydrogen Balmer lines \citep[e.g.,][]{Greene_2024}, 
indicating the presence of massive black holes (BHs).
The abundance of these AGNs in cosmic volume
\citep{Kokorev_2023,Kocevski_2024, Akins_2024} is significantly higher than what is expected from ground-based quasar surveys \citep[e.g.,][]{Matsuoka_2018,Matsuoka_2023,Niida_2020,He_2023}, 
and provides insights on the cosmological evolution of quasar luminosity functions \citep{Li_2023b}, the impact on cosmic reionization 
\citep[e.g.,][]{Dayal_2024,Madau_2024}, and the implications to rapid spins of these high-redshift BHs 
through the measurement of the radiation-to-mass conversion efficiency \citep{Inayoshi_Ichikawa_2024}.

Despite their significance, the properties of these newly identified LRDs remain puzzling, particularly concerning the origin of their 
characteristic v-shaped spectral energy distribution (SED) \citep[e.g.,][]{Barro_2023,Kocevski_2023,Labbe_2023}.
These SEDs feature a red continuum in the rest-frame optical spectrum at $\gtrsim 4000~{\rm \AA}$ and a blue (or flat) excess in the rest-frame UV range\footnote{Throughout this work, we present SEDs in the wavelength ($\lambda$) vs. flux density ($f_\nu$ or AB magnitude) plane. The characteristic v-shaped SEDs of LRDs are more apparent in the $\lambda$-$f_\lambda$ plane.}.
Observational studies using SED template fitting suggested that the red optical part could either be due to dust-reddened AGNs or starburst galaxies, 
while the blue UV component might represent light from unobscured galaxies or scattered light from embedded AGNs.

Furthermore, rest-frame near- to far-infrared observations offer additional insights.
Data from the JADES, COSMOS-Web and PRIMER-COSMOS surveys show that most LRDs are not detected in the JWST/MIRI F1800W bands,
and even for the detected sources, the MIRI SEDs tend to flatten from 
the rest-frame optical to the near-infrared (NIR) at 1--3~$\mu$m (e.g., \citealt{Williams_2024,Akins_2024}; 
see also \citealt{Wang_2024}).
Those observations appear to contradict expectations that the SED should be very red throughout the optical to NIR spectrum due to re-emission 
from hot dust components heated by the AGNs.
Consequently, these results have led to arguments disfavoring the scenario accounting for dust-obscured AGNs. 
While a certain fraction of LRDs show extremely red SEDs from NIR to mid-infrared (MIR) \citep{Lyu_2024,Perez-Gonzalez_2024}, the reasons for 
such diversity at $\lambda \gtrsim 1~\mu\mathrm{m}$ are not well understood. 
Alternatively, if considering dust-obscured galaxies instead of AGNs, cold dust associated with starbursts would dominate the far-infrared (FIR) light. 
However, current ALMA observations provide no strong evidence to support this scenario \citep{Labbe_2023,Williams_2024,Akins_2024,Casey_2024}.

It is worth noting that LRDs are extremely compact \citep[effective radius of $R_\mathrm{eff}<100$~pc at $z\sim5$;][]{Labbe_2023,Furtak_2023a} and almost point-like objects even with supreme spatial resolution by JWST.
Some red sources with a v-shaped SED show extended morphology in their images of shorter-wavelength filters, which may indicate the presence 
of an underlying host galaxy \citep{Killi_2023} or diffuse ionized gas. 
Recent work by \citet{Chen2024Morphology} reported that 50\% of LRDs in their eight LRD samples do not have detectable extended components, allowing us to place upper limits on the mass and size of the galaxies.
Comparing the BH mass estimated from broad-line emission spectroscopy to the upper limits on galaxy mass, the BH-to-galaxy mass ratio in LRDs \citep[e.g.,][]{Pacucci_2023,Kokorev_2023,Greene_2024} is 
substantially higher than the empirical value observed in the nearby universe \citep[e.g.,][]{Kormendy_Ho_2013}.
This implies the existence of an evolutionary stage where growing BHs occupy $\sim {\rm a~few}-10\%$ of the total BH+galaxy system 
\citep[for theoretical results see e.g.,][]{Inayoshi_2022, Hu_2022b,Scoggins_2023}.

Another intriguing feature of LRDs is their lack of detectable X-ray emission \citep[][]{Ananna2024,Yue2024,Maiolino2024Xray}. One possible explanation for this X-ray weakness is that the central BHs powering LRDs accrete at super-Eddington rates, where X-ray emission may be intrinsically weak (\citealt{Pacucci&Narayan2024,Madau&Haardt2024,InayoshiX_2024}).
Alternatively, the X-ray emission may be absorbed by dust-free, high column density gas in the broad-line region \citep{Maiolino2024Xray}. 
In this paper, we focus on the regions far outside the X-ray coronae and broad line regions, so that our model does not conflict with either scenario.

In this paper, we study a SED model for LRDs from rest-frame UV to infrared bands.
We hypothesize that the incident radiation from an AGN, characterized by a typical SED, is embedded in an extended dusty medium
with an extinction law similar to those seen in dense regions such as Orion Nebula or certain AGN environments.
The UV-optical spectrum is described by dust-attenuated AGN emission, composed of a red optical continuum 
and a UV excess established through gray extinction.
There is no need for additional stellar emission or AGN scattered light.
In the infrared, the SED is shaped by an extended dust and gas distribution, which allows relatively cool dust temperatures 
to dominate the radiation, thereby shifting the energy peak from near- to mid-infrared bands.
This model can produce an infrared SED flattening that is consistent with LRD observations through JWST MIRI.
We finally provide discussion and open questions rising from our conclusions regarding
(i) the temporal emergence of LRDs at $z>4$, (ii) the universality of v-shaped SEDs, and (iii)
the cosmological evolution of early BH populations.

\section{The ``v"-shape SED in rest-frame UV/optical}\label{sec:V shape}

It is evident that dust extinction curves in AGN significantly deviate from those observed in the Milky Way \citep[e.g.,][]{Maiolino_2001a,Maiolino_2001b,Maiolino_2004, Gaskell_2004,Li_2008,Xie_2017}, 
showing a characteristic flat wavelength dependence at $\lambda<4000$~\AA.
This feature is believed to stem from a deficiency of small-size grains within the AGN environment, 
which can be attributed to dust destruction mechanisms such as thermal sublimation and sputtering \citep[e.g.,][]{Laor_Draine_1993,Tazaki_2020b}, 
and grain charging \citep[e.g.,][]{Draine_Salpeter_1979}.
Recently, \cite{Tazaki_Ichikawa_2020} have applied these dust destruction processes specifically at the polar regions 
of $1-10~\pc$ directly irradiated by the AGN \citep[e.g.,][]{Honig_2017}, and show the critical role of Coulomb 
explosions in flattening the extinction curve.

\begin{figure}
\centering
\includegraphics[width=80mm]{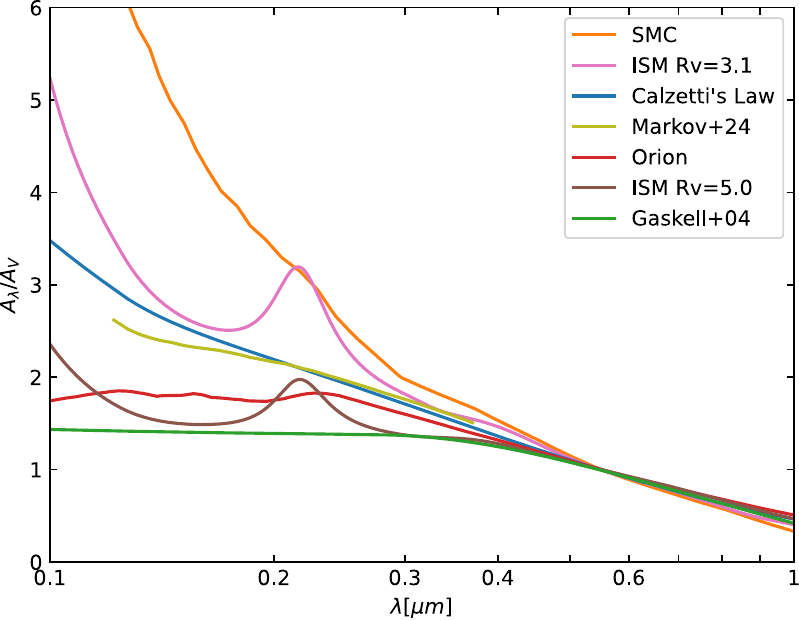}
\caption{Extinction curves as a function of wavelength normalized by the visual extinction at 5500 {$\rm \AA$}:
the Small Magellanic Cloud \citep[SMC,][]{Gordon2003}, starburst galaxies \citep{Calzetti_2000}, the interstellar dust 
in Milky Way with $R_\mathrm{V}=3.1$ and 5.0 \citep{Cardelli_1989ApJ}, high-redshift galaxies at $6<z<12$ \citep{Markov_2024}, 
the Orion Nebula \citep{Baldwin1991}, and composite AGN spectra \citep{Gaskell_2004}.
Extinction curves in the absence of small-size dust grains, as observed in the Orion Nebula,
are significantly flattened at $\lambda <2500~{\rm \AA}$, unlike the other models commonly used in galaxy SED fitting methodology \citep[e.g. Calzetii's law in][]{Labbe_2023}.
}
\label{fig:ReddeningLaw}
\end{figure}

Figure~\ref{fig:ReddeningLaw} presents seven different dust extinction laws, normalized at the visual extinction level at $5500~{\rm \AA}$ ($A_\lambda/A_\mathrm{V}$):
the Small Magellanic Cloud \citep[SMC,][]{Gordon2003}, starburst galaxies \citep{Calzetti_2000}, 
the interstellar dust in Milky Way with $R_\mathrm{V}=3.1$ and 5.0 \citep{Cardelli_1989ApJ}, 
high-redshift galaxies at $6<z<12$ \citep{Markov_2024}, the Orion Nebula \citep{Baldwin1991}, and
composite AGN spectra \citep{Gaskell_2004}.
The SMC and Calzetti's laws are commonly used in galaxy SED fitting methodology \citep[e.g.,][]{Carnall_2018,Boquien_2019}.
These models suggest that the extinction level nearly monotonically increases to shorter wavelengths.
The extinction curve for the Milky Way interstellar dust also shows a similar behavior with a bump at $\lambda \sim 2175~{\rm \AA}$,
which is considered to originate from small graphite grains and/or polycyclic aromatic hydrocarbon nanoparticles \citep{Draine_Lee_1984}.
In contrast, the extinction law measured in the Orion Nebula, one of the brightest nebulae powered by the central massive star forming region,
exhibits a curve that is significantly flattened in the far-UV wavelength range and shows the absence of the $2175~{\rm \AA}$ bump. 
The relatively grey extinction curve can be explained by the deficit of small-size dust grains. The turnover wavelength $\lambda_t$ where the extinction curve flattens is related to the minimum grain size $a_{\rm min}$ as $\lambda_t \sim 2\pi a_{\rm min}$.

Importantly, similar extinction laws are also observed in those of AGNs (\citealt{Gaskell_2004}; see also \citealt{Maiolino_2001b}, \citealt{Xie_2022}), 
which might preferentially remove small grains with sizes of $a\lesssim 0.1~\mum$. 
Additionally, JWST NIRSpec observations indicate a flattening trend of dust extinction in star-forming galaxies at $6<z<11$,
potentially due to the formation of large-size dust grains within energetic supernova ejecta \citep{Markov_2024}.

\begin{figure}
\centering
\includegraphics[width=84mm]{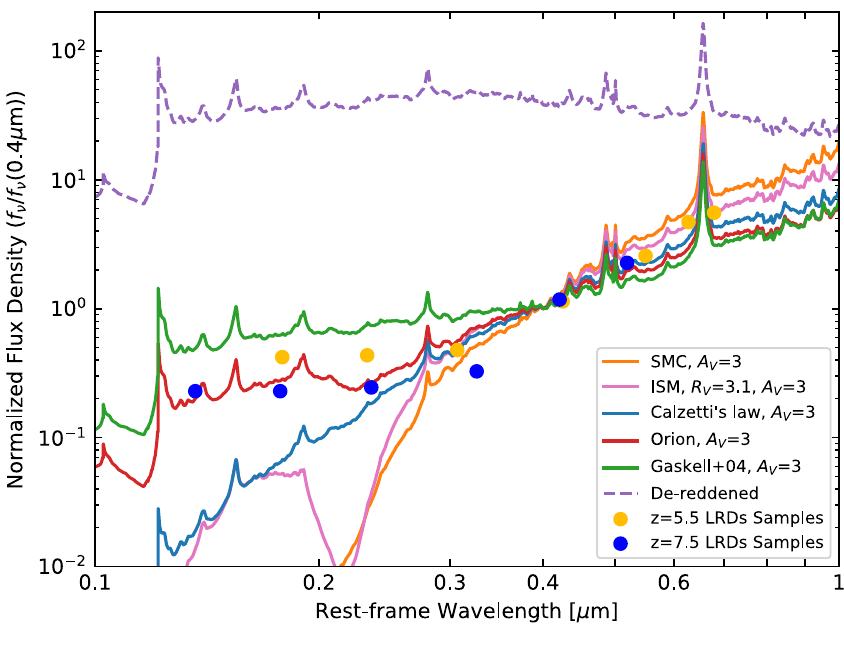}
\caption{The flux densities of dust-reddened AGNs normalized at $4000~{\rm \AA}$ (solid curves),
where the incident quasar SED is adopted from a SED template model with a removal of host galaxy and dust emission contributions 
\citep[dashed curves,][]{Temple_2021}.
The six different extinction laws shown in Figure~\ref{fig:ReddeningLaw} are applied by setting a visual extinction of $A_{V}=3$ mag. The Markov's extinction curve is not used because of the lack of full wavelength coverage.
The SED shaped with the Orion Nebula extinction law reproduces the characteristic v-shape SED of LRDs.
The averaged photometric data of LRDs at $z\sim 5.5$ and $7.5$ are overlaid for comparison \citep{Barro_2023}.
}
\label{fig:TempleDifferentAttenuation}
\end{figure}

In Figure \ref{fig:TempleDifferentAttenuation}, we show the attenuated flux using the five extinction laws.
We here adopt the composite SED of low-$z$ quasars from \citet{Temple_2021} as the incident flux.
This model provides empirical SED templates for unobscured quasars with $E(B-V)<0.3$ by using a total of $\sim 1.5\times10^4$
SDSS quasars with redshifts $0 <z< 5$. The SED
covers from the rest-frame UV to NIR up to 3 $\mum$.
In these templates, the contribution from starlight from the host galaxy is quantified using large quasar samples 
and subtracted to isolate the pure quasar SED.
For this work, we generate the SED template at $z\simeq 5$, which excludes both the host galaxy and dust contribution.
In contrast, the widely-used quasar SED template provided by \citet{VandenBerk_2001} is known to retain significant galaxy contributions,
leading to a substantially redder spectrum at $\lambda \gtrsim 5000~{\rm \AA}$ compared to those in \citet{Temple_2021}.
Although this discrepancy between the two SED models does not impact our analysis of the rest-frame UV-to-optical characteristics of LRDs,
it introduces a serious inconsistency with rest-frame NIR observations due to inadequate subtraction of host galaxy contributions 
(e.g., MIRI photometry; \citealt{Perez-Gonzalez_2024}).
Therefore, for the purpose of accurately modeling LRD SEDs in the absence of host galaxy effects, we intentionally use the model 
from \citet{Temple_2021} as the incident flux.

For all the cases, we assume a visual extinction $A_\mathrm{V}=3$ mag and normalize the attenuated fluxes at $\lambda =4000~{\rm \AA}$.
As reference, we overlay photometric SED data obtained from stacking LRD samples at $z\sim 5.5$ (orange) and $z=7.5$ (blue) \citep{Barro_2023}.
The attenuated flux density reproduces a red continuum component at $\lambda \gtrsim 3000-4000~{\rm \AA}$ with 
a slope of $\beta_{\rm opt}\gtrsim 0$, where the slope is defined as $\beta \equiv {\rm d}\ln f_\lambda /{\rm d}\ln \lambda$.
However, the SED shape significantly differs in the rest-UV part, depending on the assumed extinction curves.
Models with extinction laws observed in the Orion Nebula and AGN composite spectra flatten the SED with 
a UV slope of $\beta_{\rm UV}\sim -2$.
In contrast, other models result in strongly attenuated flux that fails to maintain a UV excess.
In the two models, the incident UV flux is reduced by two orders of magnitude, i.e.,  $1\%$ of the incident flux.

The application of the extinction law derived from AGN composite spectra \citep{Gaskell_2004}
was tested on a LRD (ID: J0647-1045) identified at $z=4.5321$ by JWST NIRSpec \citep{Killi_2023}.
For this particular object, the inferred extinction curve matches that proposed by \citet{Gaskell_2004}, although this source has an extended morphology in rest-frame UV, suggesting potential contributions from stellar components.
Our findings in this work further demonstrate that such gray extinction curves can explain the characteristic v-shaped SEDs 
of LRDs, as evidenced by accumulated large samples, without considering multiple components such as 
host galaxies or scattered AGN light.

Beyond the photometric stacking analysis \citep{Barro_2023,Noboriguchi_2023}, spectroscopic observations on LRDs reveal a universal turnover feature in the rest-frame 3000 to 4000 ${\rm \AA}$ range, with small variation among sources \citep[e.g.][]{Kocevski_2023,Greene_2024,Furtak_2024,Wang_2024,Akins_2024}. 
In our framework, this turnover wavelength $\lambda_t$ depends on the minimum dust-grain size as $\lambda_t \sim 2\pi a_{\rm min}$. While the Orion extinction curve assumes a fixed minimum dust size \citep[$a_{\rm min}\sim 0.03 ~\mum$;][]{Baldwin1991}, we propose to adopt extinction laws with different dust-size distributions to account for potential variations in the turnover point, reflecting dust destruction and growth processes in AGN environments \citep[e.g.,][]{Tazaki_Ichikawa_2020}.

Previous observational studies have proposed that the observed excess in rest-frame UV ($\beta_{\rm UV}\sim -2$)
may be attributed to the scattering of unobscured quasar light \citep{Kocevski_2023,Labbe_2023,Greene_2024}.
According to this scenario, the fraction of the scattered flux relative to the primary component heavily depends on 
both the covering factor of the scattering medium and the viewing angle, estimated to be about 1-3\% to account 
for the observed UV characteristics of LRDs.
However, this scenario does not account for the scattering fraction uniformly observed across most LRDs.
Relying on such a coincidental process to explain the universal characteristics of an abundant AGN population is challenging.
In contrast, our model, which applies the dust opacity law observed in nearby or low-redshift AGNs, offers a more robust scenario 
for the LRD characteristics (see also discussion in Section~\ref{sec:dustdistr}), 
and furthermore bring us new insights on the AGN structure within parsec scales.

\begin{figure}
\centering
\includegraphics[width=84mm]{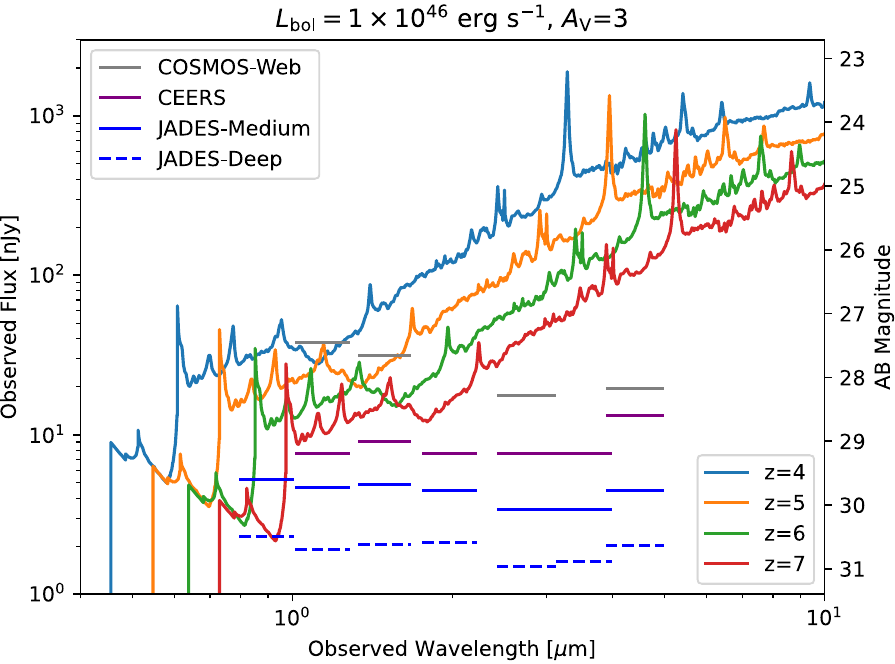}
\caption{The observed SEDs of LRDs across various redshifts of $4\leq z\leq 7$,
where the intrinsic AGN luminosity is set to $L_{\rm bol} = 10^{46}~{\rm erg~s}^{-1}$.
We adopt the composite SED template from \citet{Temple_2021}, and assume the Orion-like extinction law with $A_{\rm V}=3$ mag.
The 5$\sigma$ point-source imaging depths of JWST survey programs are overlaid: 
COSMOS-Web \citep[gray,][]{Casey_2023}, CEERS \citep[purple,][]{Finkelstein_2023}, and JADES-Medium/Deep 
\citep[blue solid and dashed,][]{Eisenstein_2023}.
}
\label{fig:TempleDifferentRedshifts}
\end{figure}

Figure~\ref{fig:TempleDifferentRedshifts} presents the SEDs of our LRD model at various redshifts from $z=4$ to $z=7$.
For these models, we assume a bolometric luminosity of $L_{\rm bol}=10^{46}~{\rm erg~s}^{-1}$ as the incident radiation from the AGN,
and calculate the monochromatic luminosity at rest-frame $1450~{\rm \AA}$ using a bolometric correction factor 
of $f_{\rm bol,UV}=4.4$ \citep{Richards_2006}.
Each case employs the Orion-Nebula attenuation law with a visual extinction $A_\mathrm{V}=3~{\rm mag}$, as shown by the red curve in Figure~\ref{fig:TempleDifferentAttenuation}.
For comparison, we overlay the $5~\sigma$ point-source imaging depths for each NIRCam broadband filter used in several JWST survey programs.
The depths achieved by the CEERS and JADES surveys are sufficient to detect the characteristic SED shape of LRDs from the 
rest-frame UV to optical wavelengths.
This is consistent with the observational fact that LRDs identified in these deep surveys exhibit bolometric luminosities similar to our models, 
$L_{\rm bol}\sim 10^{45-46}~{\rm erg~s}^{-1}$ \citep[e.g.,][]{Kokorev_2024}. 
However, the COSMOS-Web survey, which covers wider areas ($\sim 0.5$~deg$^2$, approximately $\sim20$ times larger than other surveys 
with $\sim100$~arcmin$^2$) with relatively shallower depths (as shown in Figure~\ref{fig:TempleDifferentRedshifts}), requires higher bolometric luminosities
for the UV component to be detectable and to capture the characteristic SED shape.
Our SED models satisfy the selection criteria for LRDs; $\beta_{\rm UV} < -0.37$ and $\beta_{\rm opt}>0$ (\citealt{Kocevski_2024}; see also \citealt{Greene_2024}).

\section{Reprocessed Infrared Radiation from Obscuring Dust} \label{section:IR}

Next, we analyze the SED of LRDs in the IR band, focusing on the re-emission from dust grains that obscure
the AGN light from UV to optical wavelengths.
Since UV photons emitted from the AGN heat the obscuring material, the dust temperature rises to a sublimation point of
$T_{\rm sub}\simeq 1500~\K$, and decreases with distance from the AGN.
In scenarios where dust and gas are centrally concentrated, as in dusty tori \citep[e.g.,][]{RamosAlmeida_2017}, the hot dust just 
behind the sublimation radius significantly contributes to the dust-grain mass budget and thus preferentially re-emits 
thermal heat into the NIR bands, $\lambda \simeq 2.9~\mum~(T_{\rm dust}/10^3~\K)^{-1}$.
As a result, the NIR flux density can be $\gtrsim 2$ dex higher than the attenuated AGN emission at the optical band,
resulting in an extremely red continuum from the optical to infrared regimes, as observed in dust-obscured AGN
at lower redshifts 
\citep[e.g.,][]{AlonsoHerrero_2006,Dey_2008,RamosAlmeida_2011,Ichikawa_2015,Ichikawa_2017,Toba_2017,Lyu_2022}.
However, a good fraction of high-redshift LRDs observed with MIRI broadband filters (e.g., F770W and F1800W) do not show 
such red color but rather exhibit a relatively flat SED from the rest-frame optical to the NIR wavelengths \citep{Williams_2024,Wang_2024}.
This trend is also seen in stacking images of multiple LRDs \citep{Akins_2024}.
Nevertheless, there remain LRDs that show extremely red SEDs from NIR to MIR \citep{Lyu_2024, Perez-Gonzalez_2024}.

The shape of the IR SED, reprocessed by dust grains heated by an AGN, depends on the radial distribution of the dust
\citep[e.g.,][]{Pier_1992,Nenkova_2008,Nikutta_2021}.
When the obscuring material does not form a concentrated density structure or is dispersed by bipolar outflows, 
the peak energy shifts to the MIR since the bulk of the re-emission comes from relatively cooler dust located farther 
from the dust sublimation radius \citep[e.g.,][]{Barvainis_1987}. 
NIR interferometry observations of nearby AGNs in recent decades \citep[e.g.,][]{Kishimoto_2011b,Kishimoto_2011} have found that 
the NIR emitting region appears to have a radius nearly twice as large as the expected sublimation radius, as traced by the NIR dust reverberation mapping observations \citep{Suganuma_2006,Koshida_2014}.
This suggests that the dust density profile has a slightly lower level of concentration, producing a larger emissivity size than the sublimation radius \citep[see also][]{Nikutta_2021b}.
Additionally, MIR interferometry observations resolving the central $\sim$pc scale also showed that the significant fraction of MIR dust emission originates from the polar region, suggesting the presence of bipolar dusty outflows
\citep[e.g.,][]{Jaffe_2004, Hoenig_2012,Hoenig_2013,Hoenig_2019,Tristram_2014}.
According to radiation transfer calculations that explain the IR SED of AGNs, including dust emission towards the polar region, 
the gas and dust density profiles with a low level of concentration ($\rho \propto r^{-\gamma}$;
$-0.5<\gamma<0.5$) can reproduce the observed MIR bump along with a relatively flat NIR SED at $\lambda \simeq 3-5~\mum$ \citep{Honig_2017}.

\begin{figure*}
\centering
\includegraphics[width=85mm]{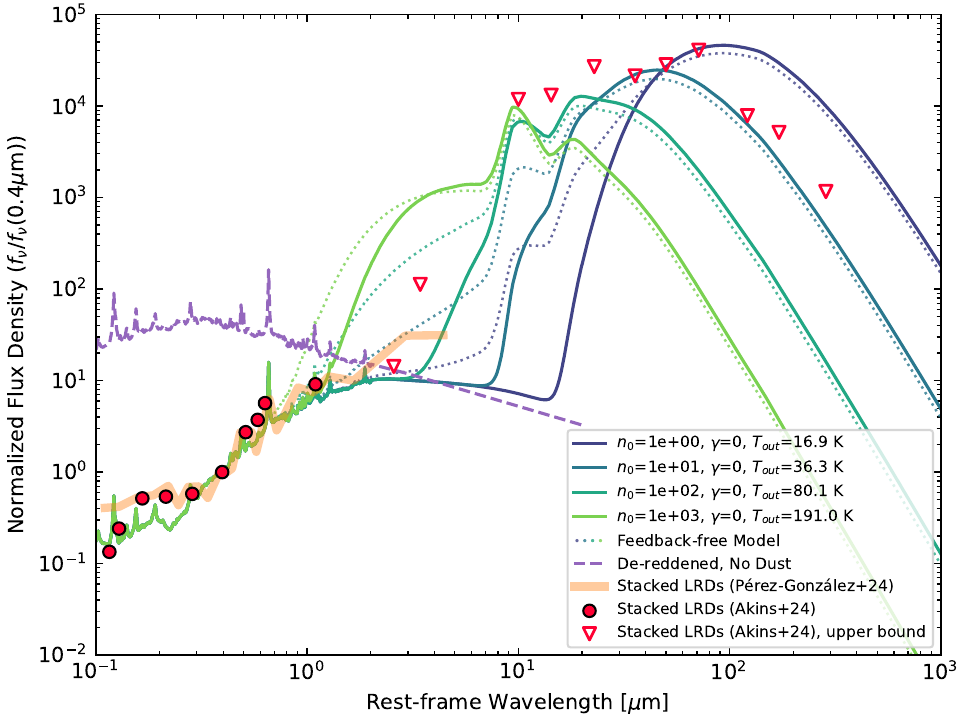}\hspace{6mm}
\includegraphics[width=85mm]{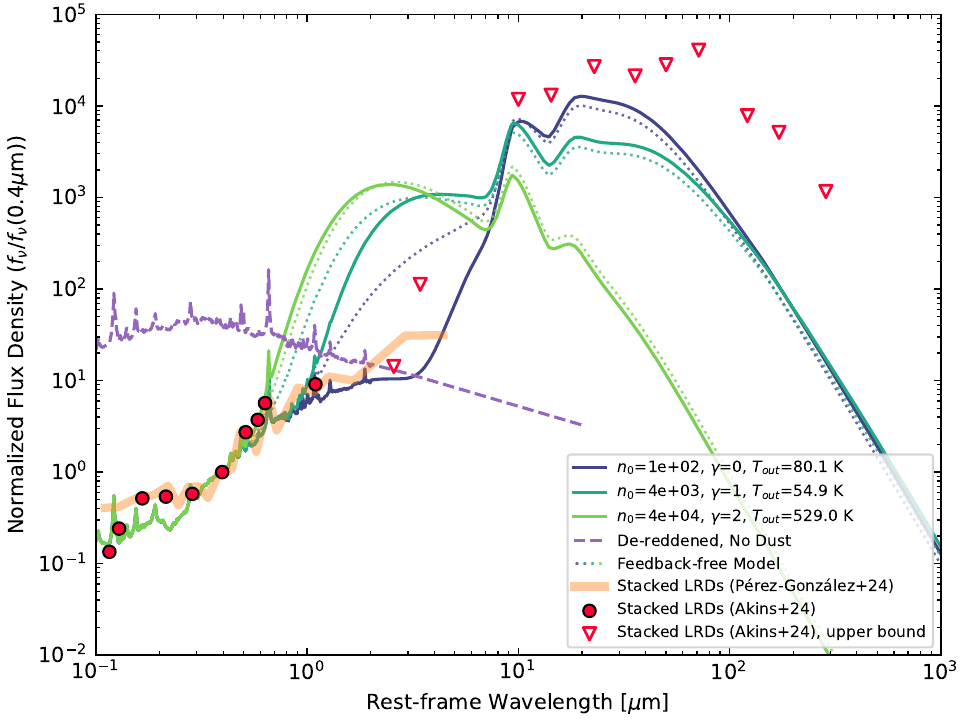}
\caption{The flux densities of LRDs normalized at $4000~{\rm \AA}$ for different values of $n_0$ and $\gamma$, 
where the incident SED is the same as in Figure~\ref{fig:TempleDifferentAttenuation}. 
The stacked photometric data of LRDs are taken from the JADES survey \citep[thick curve,][]{Perez-Gonzalez_2024}
and the COSMOS-Web survey \citep[circles][]{Akins_2024}, and the upper bounds of the mid- to far-IR flux densities 
due to non-detection in SCUBA and ALMA bands are taken from \citet{Akins_2024}.
Left panel: dependence of the SED on the density $n_0$.
The highest density case of $n_0=10^3~\cc$ results in overly bright rest-frame NIR, violating the MIRI F770W constraints.
Lower density cases reproduce SEDs consistent with the observed data points, while the lowest density of $n_0=1~\cc$ yields
FIR SED peak inconsistent with the upper bounds from the ALMA observations.
Right panel: Variation of SEDs with different density slopes $\gamma$.
As the density is less concentrated ($\gamma<1$), the energy peak shifts from the NIR to the MIR regime.
The feedback-free model for each case is presented with the dotted curve.
}
\vspace{3mm}
\label{fig:IR_SED}
\end{figure*}

Motivated by these AGN IR observations and models, we focus not only on a compact dust structure such as a dust torus, 
but rather a more extended dust structure described by a radial density profile of $\rho = \rho_0 (r/r_{\rm in})^{-\gamma}$, 
where $\rho_0$ is the gas density at the sublimation radius $r_{\rm in}$, and the power-law index ranges $0<\gamma<2$. 
With the underlying density distribution, the temperature of dust grains, $T_{\rm dust}(r)$, 
can be determined by balancing dust heating from irradiation with cooling through thermal emission as
\begin{equation}
\int \frac{L_{\nu} e^{-\tau_\nu(r)}}{4 \pi r^2}\kappa_{\nu}^{\rm abs} d\nu =
4 \pi \int \kappa_{\nu}^{\rm abs} B_\nu\left[T_{\mathrm{dust}}(r)\right] d \nu,
\label{eq:energy conservation}
\end{equation}
where $L_{\nu}$ is the luminosity density of the central AGN at a frequency of $\nu$, $\kappa_{\nu}^{\rm abs}$ is the absorption opacity of dust, 
$B_\nu(T_{\rm dust})$ is the Planck function with a dust temperature $T_{\rm dust}(r)$ at a distance of $r$ from the center,
and $\tau_{\mathrm{\nu}}$ is the optical depth defined by
\begin{equation}
\tau_\nu (r)= \int_{r_{\rm in}}^{r} \rho \kappa_{\nu}^{\rm ext} dr',
\end{equation}
where $\kappa_{\nu}^{\rm ext}$ is the sum of the absorption and scattering opacity.

With a dust temperature profile, the luminosity density of dust re-emission is calculated by
\begin{equation}
L_{\nu, \mathrm{dust}}=4\pi \Omega \int_{r_{\rm in}}^{r_{\rm out}} \kappa_{\nu}^{\rm abs} B_\nu \left[T_{\mathrm{dust}}(r)\right] \rho {r}^2 d r,
\label{eq:IRdust}
\end{equation}
where $0<\Omega \leq 4 \pi$ is the solid angle covered by dust and $r_{\rm out}$ is determined by a constraint on 
the column density so that the visual extinction reaches $A_{\rm V}\simeq 3$ mag to reproduce the characteristic red continuum spectra
of LRDs in the rest-frame optical bands (see Section~\ref{sec:V shape}).
In our calculation, we assume a full-sky covering case (i.e., $C\equiv \Omega/4\pi=1$), while this value 
will be determined when the relative cosmic abundance between LRDs and unobscured AGNs and the dependence of $C$ on 
the AGN properties such as the Eddington ratio are better understood for LRDs \citep[e.g.,][]{Ichikawa_2019,Ricci_2017,Ricci_2023}.
Note that the equation~(\ref{eq:IRdust}) can be applied in the optically thin limit of dust re-emission in the dusty surrounding medium.
This approximation is valid for a wide range of parameter combinations of the density value ($n_0\equiv \rho_0/m_{\rm p}$) and 
slope ($\gamma$) of the distribution we explore in our SED modeling for LRDs (see Section~\ref{sec:discussionIR}).

In addition to a diluted density distribution, we consider the feedback effect on the dusty medium surrounding the central AGN
\citep[e.g.,][]{Chan_2016,Namekata_2016,Kudoh_2023,Soliman_2023}. 
Radiation pressure expels dust grains between the sublimation radius and the dust photosphere to UV radiation into
a thin shell that accumulates at $r=r_{\rm ph}$, the radius of the photosphere where the optical depth measured 
from the center reaches unity. 
Within the thin layer at $r=r_{\rm ph}$, the optical depth transitions from 0 to 1, and the dust temperature varies 
depending on the detailed density distribution. 
For simplicity, we assume a constant temperature within this layer, corresponding to the inner edge of the thin shell.
This temperature can be determined using Equation~\eqref{eq:energy conservation} with $r = r_{\rm ph}$ and $\tau_\nu(r) = 0$, 
setting an upper bound on the temperature and emission from the thin shell. 
In our model, feedback effects are incorporated by substituting $r_{\rm in}$ with $r_{\rm ph}$ in Equation~(\ref{eq:IRdust}).
While detailed radiation hydrodynamic simulations can pinpoint the exact location of the photosphere,
we evaluate $r_{\rm ph}$ using the original unperturbed density profile, because the optical depth value 
within a radius of $r$ remains unchanged if mass is conserved.
As a result of the feedback, hot dust just behind the sublimation radius $r_{\rm in}$ is pushed to larger radii and becomes cooler, 
shifting the NIR emission to MIR.
Despite these change within $r_{\rm ph}$, the temperature profile and emission of the region outside $r_{\rm ph}$ remain unaffected.

Figure~\ref{fig:IR_SED} presents multi-wavelength SEDs of a LRD from the rest-frame UV to far infrared with various combinations of 
($n_0$, $\gamma$).
This wavelength coverage corresponds to the observed wavelengths frame from JWST NIRCam/MIRI to ALMA bands for sources at $z=6$. 
For comparison, we overlay the stacking photometric data of observed LRDs reported by the JADES survey 
\citep[thick curve,][]{Perez-Gonzalez_2024} and the COSMOS-Web survey \citep[circles,][]{Akins_2024}. 
Furthermore, the upper bounds of the mid- to far-IR flux densities 
due to non-detection in SCUBA and ALMA bands are shown (triangles, \citealt{Akins_2024}; see also \citealt{Labbe_2023}).

In the left panel of Figure~\ref{fig:IR_SED}, we show the dependence of the SED shape on the gas density ($1\leq n_0/\cc \leq 10^3$) 
at the sublimation radius, while the slope of its radial profile is fixed to $\gamma = 0$. 
When the density is as high as $n_0=10^3~\cc$, the outer edge of the obscuring material is located at $r_{\rm out}\gtrsim 30~r_{\rm in}$, 
where the dust temperature ranges from $T_{\rm dust}\simeq 1500~\K$ to $\sim 200~\K$.
Therefore, the reprocessed IR spectrum begins to rise from $\lambda \simeq 1~\mum$ and exhibits a near- to mid-infrared bump 
that covers $\lambda \simeq 2-15~\mum$.
In this case, the bright NIR bump makes the SED shape substantially redder from $\lambda \simeq 6000~{\rm \AA}$ to $\sim 2~\mum$, consistently 
observed in AGNs with hot dust tori \citep{Polletta_2007}.
However, this extremely red SED color does not match the JWST MIRI photometric data for LRDs, which suggest a modest red color or even a flat spectral shape. 
We note that such a contradiction with the MIRI data appears as long as $n_0 \gtrsim 10^3~\cc$.

\begin{figure}
\centering
\includegraphics[width=85mm]{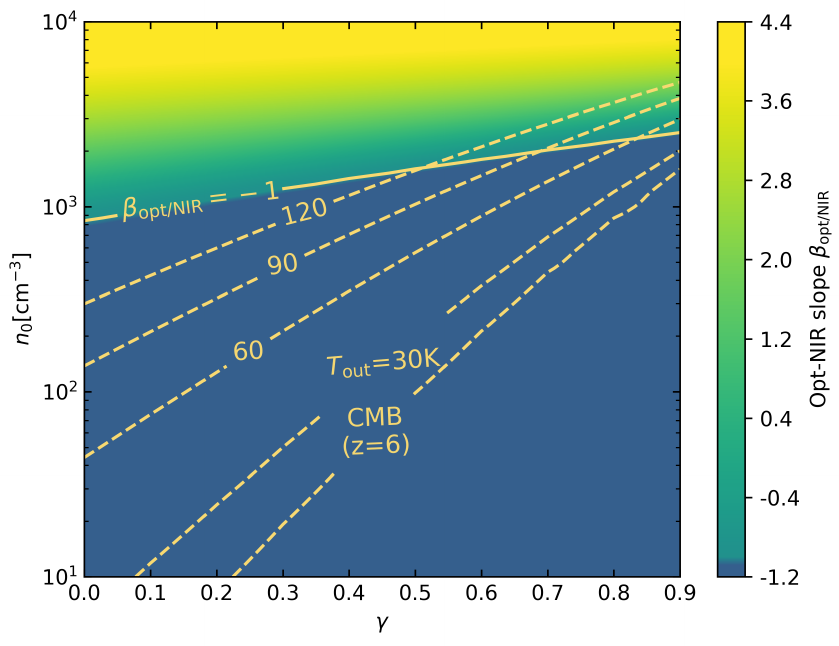}
\caption{
Opt-NIR slope $\beta_\mathrm{opt/NIR}$ as the function of density power-law index $\gamma$ and gas density $n_0$. 
High density cases lead to bright rest-frame NIR emission and thus a steeper slope.
The solid curve shows the boundary of $\beta_\mathrm{opt/NIR}=-1$, below which the NIR SED in our model is consistent 
wit the requirement from JWST/MIRI observations for LRDs.
The dashed curves present constant $30\leq T_\mathrm{out}/\K \leq 120$ and the temperature of Cosmic Microwave Background (CMB) at $z=6$.
Below $T_{\rm out}\simeq 30~\K$, the SED becomes inconsistent with the non-detection of LRDs with ALMA observations. 
The allowed parameter space is approximated as $10^{2.9\gamma+0.83} \lesssim n_0/\cc \lesssim 10^3$.}
\label{fig:Boundary}
\end{figure}

On the other hand, as the density $n_0$ decreases and the outer radius expands to reach a sufficient attenuation level, 
the peak energy of the IR SED decreases 
because relatively cooler dust grains dominate both the mass and IR emissivity.
The energy transfer from NIR to MIR significantly suppresses the NIR bump and eases the mismatch between our SED model and the JWST MIRI photometric data. 
For the cases with $n_0=10$ and $100~\cc$, the SED shapes from the optical band to $\lambda = 2-3~\mum$ agree to the observed spectra of LRDs (or their stacked photometry based on the F444W and F770W filters for $z\simeq 6$ LRDs), and also the peak energy is shifted to the FIR band at $\lambda \sim 20-50~\mum$.
As the density decreases to $n_0=1~\cc$, the IR SED peaks at $\lambda \sim 100~\mum$ and exceeds the upper bounds 
obtained by ALMA observations.

We also observe that our SED model with $\gamma=0$ and $n_0<10^3~\cc$ aligns with the upper bounds of the F1800W photometry.
It is important to note that the F1800W upper limit is derived from stacking images covering less than 20\% of the LRD samples observed with the F770W MIRI filter \citep{Akins_2024}. 
Further investigations into the rest-frame MIR band, such as through F1800W, are crucial for enhancing our understanding of the nature of LRDs, e.g., the PRIMER \citep{PRIMER_2021}, MEGA \citep{MEGA_survey}, MEOW \citep{MEOW_survey}, and COSMOS-3D \citep{COSMOS-3D} survey programs.

The right panel of Figure~\ref{fig:IR_SED} presents cases for a centrally concentrated density distribution with 
$\gamma=2$ and $\gamma=1$, alongside the case of $\gamma=0$ and $n_0=100~\cc$ (see the left panel).
At the sublimation radius, the densities are set to $n_0=4\times 10^4~\cc$ and $4\times 10^3~\cc$ for $\gamma=2$ and $1$, 
respectively, ensuring sufficient extinction ($A_\mathrm{V}\simeq 3$ mag).
When the density is centrally concentrated with $\gamma=1-2$, a significant fraction of the obscuring material 
is efficiently heated by the AGN, and thus the hot dust just behind the sublimation radius is responsible for the IR emissivity.
As a result, the spectrum appears redder in the NIR wavelengths, similar to the high-density case shown in the left panel.

In each panel, we show the SEDs for cases without radiation feedback effects (dotted curves). 
The feedback-free model predict brighter NIR emission compared to the models with feedback,
because without feedback the hot dust layers at the sublimation radius contributes significantly 
to the NIR emission.

\section{Discussion}
\subsection{Physical properties of gas obscuring LRDs}\label{sec:discussionIR}

In this section, we explore the parameter regions $(n_0,\gamma)$ that satisfy the conditions required for capturing the characteristic SED shape of LRDs.
We consider the following three key conditions.
\begin{enumerate}
\item Dust attenuation:
To reproduce the rest-frame UV-to-optical SED of LRDs, the dust attenuation is set to $A_{\rm V} \simeq 3$ mag. 
This level of attenuation corresponds to a hydrogen column density of $N_{\rm H}\simeq 7.5\times 10^{22}~{\rm cm}^{-2}~(Z/0.1~\zsun)^{-1}$ \citep[][]{Baldwin1991}.

\item Optical-NIR SED slope: 
The spectral slope between the rest-frame optical and NIR regions needs to be consistent with the JWST NIRCam/MIRI photometry, 
showing that the F444W$-$F770W colors for $z\simeq 5-6$ LRDs are $\simeq 0.5$ mag in stacking images.
To quantify the optical-to-NIR slope, we analyze flux densities at $\lambda = 0.65$ and $1.2~\mum$ from our SED model. 
An crucial consideration is the influence of ${\rm H}\alpha$ line emission, which can cause an excess in the NIRCam filter-convolved 
flux density at $\lambda \simeq 0.65~\mum$.
For LRD samples identified by slitless spectroscopy \citep{Matthee_2024}, we adopt an average rest-frame equivalent width (EW) of 
${\rm EW}_{0,\rm H\alpha}=500~{\rm \AA}$. 
The prominent line emission enhances the filter-convolved flux densities within a filter with a bandwidth of $\Delta \lambda_{\rm filter}$ 
by a factor of $\ell \equiv [{\rm EW}_{0,\rm H\alpha}(1+z) + \Delta \lambda_{\rm filter}]/\Delta \lambda_{\rm filter}$. 
Assuming that the ${\rm H}\alpha$ line emission falls within the F444W, this excess factor of $\ell \simeq 1.35$ applies for a $z=6$, 
corresponding to an additional $0.33$ mag.
Therefore, we define the optical-to-NIR slope as 
\begin{equation}
\beta_{\rm opt/NIR} = \frac{\log_{10}[f_{\nu_2}/(\ell f_{\nu_1})]}{\log_{10}(\lambda_2/\lambda_1)}-2,
\label{eq:IRslope}
\end{equation}
where $\lambda_1=0.65~\mum$, $\lambda_2=1.2~\mum$ to measure the continuum flux $f_{\nu}$, and $\nu_{1(2)}=c/\lambda_{1(2)}$. Here we impose 
$\beta_{\rm opt/NIR}< -1.0$ for this study.

\item ALMA non-detection:
Given the non-detection of LRDs in ALMA observations, we impose a lower limit on the dust energy that re-emits the AGN light. 
This requires that the minimum dust temperature, which is analytically calculated in our SED model, exceeds a temperature 
threshold of $T_{\rm min}= 30~\K$.

\end{enumerate}

\begin{figure*}
    \centering
    \includegraphics[width=160mm]{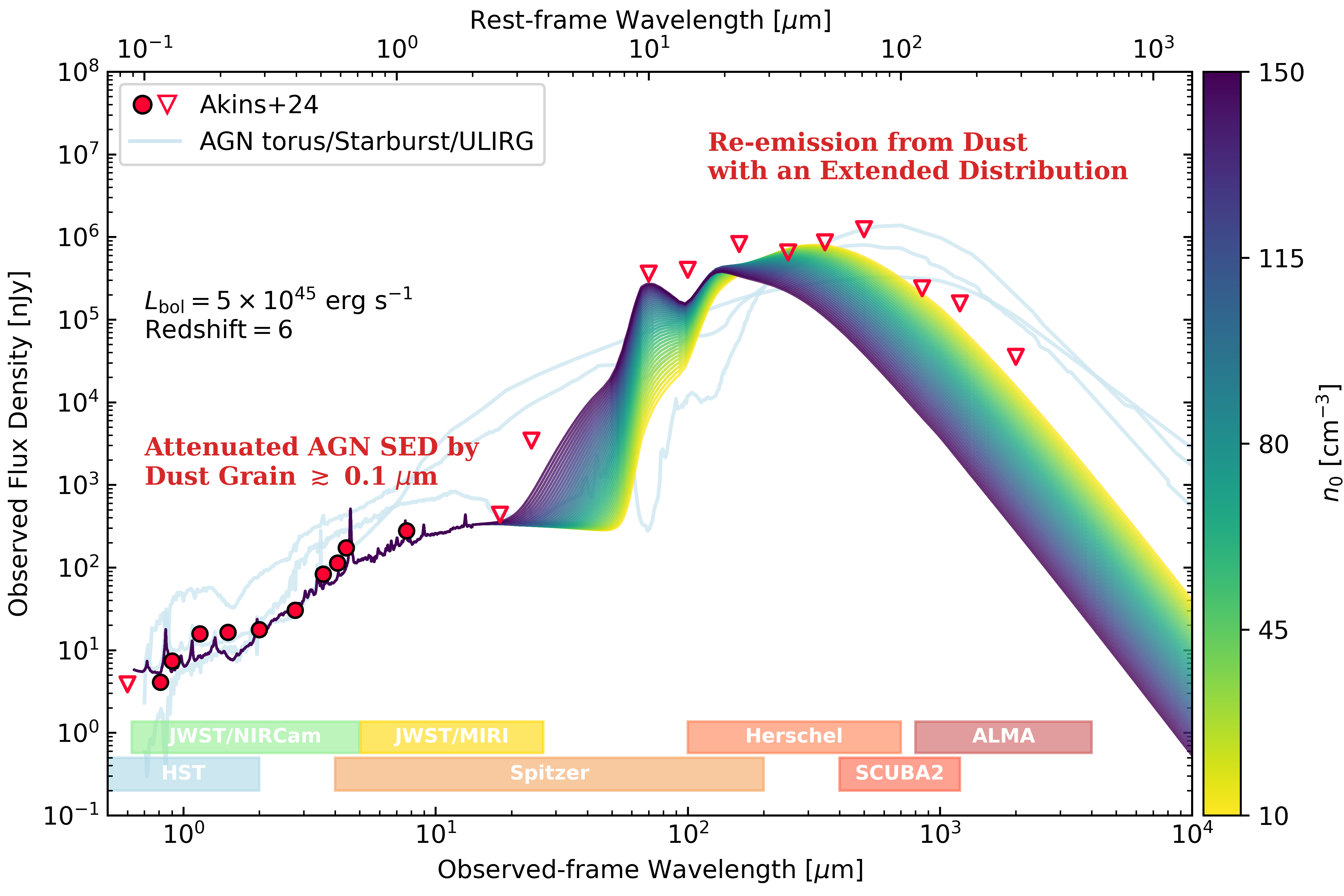}
    \caption{Observed SEDs of LRDs in our model assuming an AGN bolometric luminosity of 
    $L_\mathrm{bol}=5\times10^{45}$~erg s$^{-1}$ at $z=6$. 
    The colored curves presents variations in SEDs with gas density $n_0$, ranging from $150~\cc$ (dark color) 
    to $10~\cc$ (light color), while maintaining a constant radial distribution index $\gamma = 0$. 
    The stacked photometric data for LRDs and upper bounds talen from \cite{Akins_2024} are overlaid.
    For comparison, we also show the SED templates of AGN hot dust tori (Torus model), Apr 220, and Mrk 231
    provided by \citep[light blue curves,][]{Polletta_2007}, for which the same bolometric luminosity is assumed.
    At the bottom of the plot, we show the wavelength coverage of related telescopes.}
    \label{fig:VIP}
    \vspace{3mm}
\end{figure*}

In the analysis described in Section~\ref{section:IR}, we find that the SED peaks in the NIR regime and fails to meet 
the slope criterion of $\beta_{\rm opt/NIR}<-1$ when a centrally concentrated density profile with $\gamma > 1$ is assumed 
(see the right panel of Figure~\ref{fig:IR_SED}). 
We thus focus on the parameter region with $\gamma \leq 1$ in the following discussion.

Figure~\ref{fig:Boundary} summarizes the characteristic properties of the SED across various combinations of $n_0$ and $\gamma$.
The color contours indicate the optical-to-NIR slope as defined by Equation~(\ref{eq:IRslope}).
Employing the criterion of $\beta_{\rm opt/NIR}=-1$, the parameter space is divided into two regions.
In the high density region ($n_0\gtrsim 10^3~\cc$), the NIR SED is influenced by re-emission from hot dust grains 
heated by the AGN, resulting in a steeper optical-to-NIR slope that is inconsistent with the observational requirement.
We also present curves showing a constant dust temperature $T_{\rm out}$ at the outer edge of the obscuring material at $r=r_{\rm out}$.
In cases with lower density, the outer radius further extends to achieve $A_\mathrm{V}\simeq 3$ mag and $T_{\rm out}$ also decreases.
By imposing $T_{\rm out}\geq T_{\rm min}=30~\K$, we ensure that the FIR part of the SED aligns with non-detection
of LRDs with ALMA.

In summary, the region between the two curves of $\beta_{\rm opt/NIR}=-1$ and $T_{\rm out}=30~\K$ defines 
the allowed parameter space of $(n_0,\gamma)$. 
Here, the density range is approximated as 
\begin{equation}\label{Eq:n0gammaconstraints}
10^{2.9\gamma+0.83} \lesssim  n_0/(\cc)  \lesssim 10^3,
\end{equation}
which is valid $\gamma<0.8$.
We also note that the density values are the lower side of the typical electron density 
of the low redshift narrow line region (NLR) with $n_{\rm e}\sim 10-10^6~\cc$ \citep[e.g.,][]{Netzer_1990,Nesvadba_2006,Liu_2013,Harrison_2014}. 
Recently, Integral Field Unit (IFU) observations that spatially resolves NLRs of low-redshift AGNs show that 
the electron density at the inner $\sim 100~\pc$ scales is found to be $n_{\rm e}\sim 10^2-10^3~\cc$, and 
decreases toward larger radii \citep[][]{Kakkad_2018}.  
For high-redshift star forming galaxies at $z\sim 6$ observed by JWST/NIRSpec, the electron density in the nebular region 
is estimated as $n_{\rm e} \gtrsim 10^2~\cc$ \citep[e.g.][]{Isobe_2023,Abdurro'uf_2024}.

Finally, Figure~\ref{fig:VIP} shows the compiled SED models that agree to the MIRI (F770W and F1800W) and ALMA constraints.
We consider a LRD at $z=6$ with a bolometric luminosity of $L_\mathrm{bol}=5\times10^{45}$~erg s$^{-1}$.
The color code presents variations of gas density $n_0$, while maintaining a constant radial distribution index $\gamma = 0$. 
For comparison, we also show the SED templates of AGN hot dust tori (Torus model), Apr 220, and Mrk 231
provided by \citep[light blue curves][]{Polletta_2007}, for which the same bolometric luminosity is assumed.
The SED shapes of these compared templates violate observational constraints either at the NIR or FIR regimes.

With the median value of the density in the log space ($n_0\sim 100~\cc$),
the dust mass enclosed within $r\leq r_{\rm out}$ is estimated as $M_{\rm dust}\simeq 1.2~(9.6)\times 10^4~\msun $ for 
$\gamma=0$ and $0.2$, respectively, assuming a dust-to-gas mass ratio of $0.001(Z/0.1~\zsun)$.
These results agree well to the mean value of the dust mass $\langle M_{\rm dust}\rangle \simeq 1.6^{+4.8}_{-0.9}\times 10^4~\msun$
estimated using NIRCam images of LRDs \citep{Casey_2024}.
Adopting the dust-to-stellar mass ratio of $f_{\star,\rm dust}\sim 10^{-4}-10^{-3}$ in a galaxy younger than one billion years
\citep[e.g.,][]{Valiante_2009}, the expected mass of stars that form these dust grains would be the order of $\sim 10^{7-9}~\msun$.
This estimate can be used in comparison to other independent stellar-mass measurements, e.g., by image decompositions and SED fitting.

\begin{figure*}
\centering
\includegraphics[width=85mm]{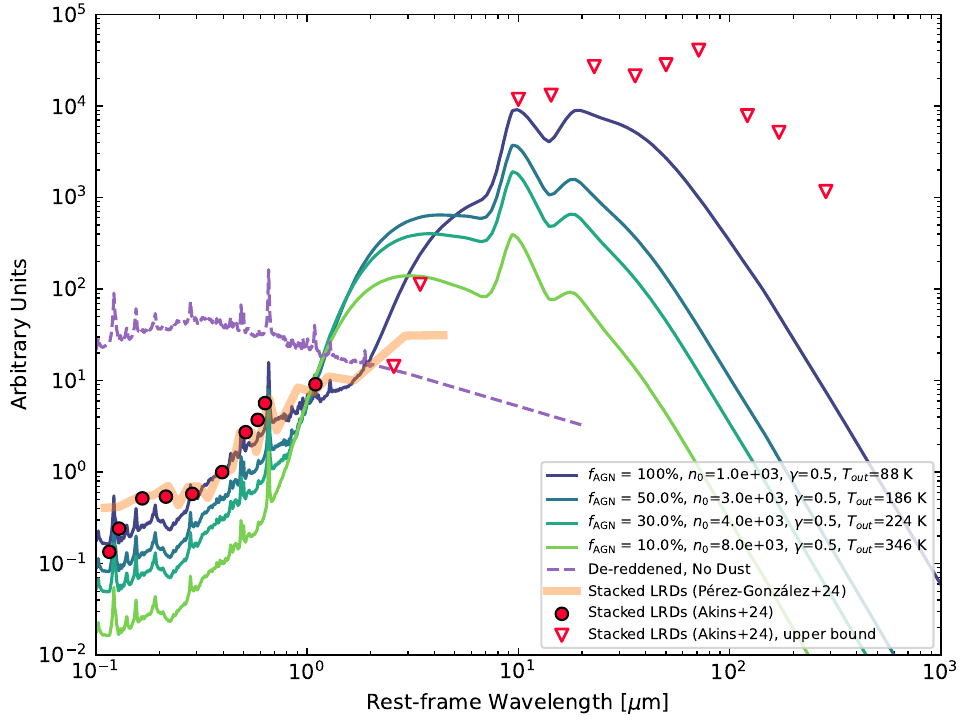}\hspace{6mm}
\includegraphics[width=85mm]{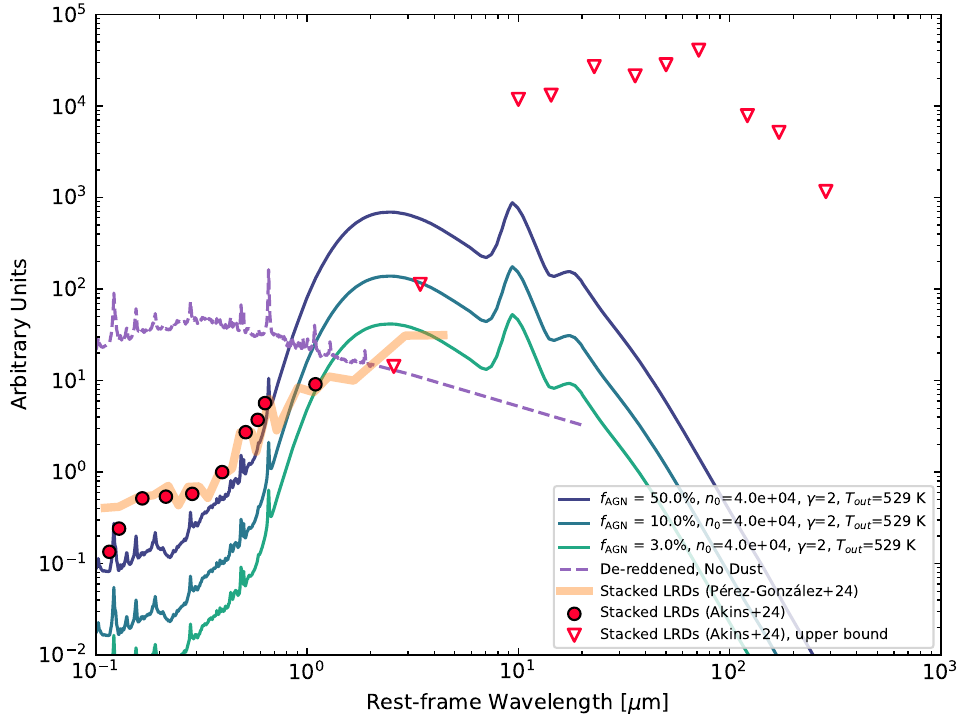}
\caption{The effect of the AGN contribution to the observed LRD SEDs. 
The fraction of $f_{\rm AGN}=100\%$ is defined such that the model SED matches 
the observation at $4000~{\rm \AA}$, with the SED normalization scaling to $f_{\rm AGN}$.
Left panel: $f_{\rm AGN}$ varies from 100\% to 10\%, with $n_0$ adjusted to match the F770W flux at rest-frame $\sim$ 1 $\mu$m. 
A lower value of $f_{\rm AGN}$ allows $n_0$ to increase from $10^3~\cc$ to $8\times10^3\ \cc$. 
Right panel: For $\gamma = 2$, representing a typical dust torus model, the highly concentrated 
dust distribution produces bright NIR emission. 
The model aligns with the MIRI photometric result only when $f_{\rm AGN} \leq 3\%$.
}
\vspace{3mm}
\label{fig:AGN_Fraction}
\end{figure*}

\subsection{Host galaxy contributions?}
Previous observations of LRDs have provided limited information about their host galaxies.
While some LRDs exhibit extended morphology in short-wavelength images, achieving a robust confirmation of 
host galaxy features remains challenging.
Detailed measurements of JWST/PRISM spectra have successfully identified a continuum break feature at 
$\lambda \simeq 4000~{\rm \AA}$ in several LRDs: MSAID13123 at $z=7.04$ \citep{Greene_2024}, RUBIES-BLAGN-1 at $z=3.1$ \citep{Wang_2024}, RUBIES-EGS-49140 at $z=6.68$, RUBIES-EGS-55604 at $z=6.98$, RUBIES-EGS-966323 at $z=8.35$ \citep{WangBingjie2024b}, and GN-72127 at $z=4.13$ \citep{Kokorev2024BalmerBreak}.
These findings suggest the presence of an underlying host galaxy.
However, if stellar light accounted for the continuum flux across the entire optical band,
the required stellar mass would approach $M_\star \gtrsim 10^{11}~\msun$, comparable to that of the Milky Way.
When combining such a high mass and their high abundance in the cosmic volume, this scenario implies an extraordinary stellar mass density 
that contradicts the standard cosmological framework \citep[see also][]{Inayoshi_Ichikawa_2024,Akins_2024}.
As a possible solution, \citet{Wang_2024} suggest that the stellar contribution on the SED is limited to 
the short-wavelength range at $\lambda \lesssim 4000~{\rm \AA}$ and the AGN is responsible for the red optical continuum
at the longer wavelengths.
Under this assumption, the SED fitting indicates a minimal stellar mass of $M_\star \sim 10^{9}~\msun$,
resolving the tension with the cosmological prediction.

By accounting for contributions from the host galaxies to the rest-frame optical and near-infrared continuum emission, the constraints on the density normalization $n_0$ and slope $\gamma$ of dusty flows can be relaxed. 
To incorporate AGN flux contributions into the observed LRD SEDs, we define $f_{\rm AGN}=100\%$, where the flux density is normalized at 4000~{\rm \AA} to match the observations. For lower $f_{\rm AGN}$ values, the modeled flux density is scaled linearly.
The left panel of Figure~\ref{fig:AGN_Fraction} shows the SEDs modeled with $\gamma=0.5$ for various AGN flux contributions from $f_{\rm AGN}=10\%$ to $100\%$. In these models, the density at the sublimation radius is adjusted to ensure that the modeled flux densities are consistent with the F770W flux density. As the AGN contribution decreases to $f_{\rm AGN}=10\%$, the maximum density value increases from $n_0\simeq 10^3~\cc$ to $8\times 10^3~\cc$ (see Figure~\ref{fig:Boundary}).
When the F1800W flux upper limit is taken into account, the maximum density value settles down around $n_0\simeq 6\times 10^2~(1.2\times 10^3)~\cc$ for $f_{\rm AGN}=100\%$ (and $10\%$, respectively).

The lower bound of the density normalization is set by the constraint on the dust temperature at the outer edge, requiring $T_{\rm out}\geq T_{\rm min}$. To ensure consistency with the ALMA non-detection, the minimum density is given as $n_0\simeq 1.9\times 10^2~\cc$ for $\gamma=0.5$ (see Equation~\ref{Eq:n0gammaconstraints}).
For $f_{\rm AGN}=30\%$, the minimum dust temperature $T_{\rm min}$ decreases to $\sim 19~\K$ when the density is set to $n_0=10^2~\cc$. However, this temperature threshold reaches the CMB temperature, suggesting that the lower bound of the density normalization is not relaxed further.

Moreover, the concentrated dusty torus model with high temperatures appears less plausible, even when accounting for host galaxy contributions. 
The right panel of Figure~\ref{fig:AGN_Fraction} shows the SEDs for a typical torus parameter ($\gamma \sim 2$) at AGN fractions of $f_{\rm AGN}=3$, $10$, and $50\%$. 
In these cases with $\gamma=2$, the density normalization is nearly fixed at $n_0=4\times 10^4~\cc$ due to the column density constraint ($A_{\rm V}=3$ mag).
To remain consistent with observational constraints, specifically the F770W flux density, the AGN fraction must be significantly reduced to $f_{\rm AGN}=3\%$ due to the strong NIR emission predicted by this torus model. 
This finding agrees with the argument in \citet{Wang_2024}, who reported that observed rest-frame NIR flux is at most $5\%$ of the predictions from standard torus models. However, such a low AGN contribution implies an intrinsic EW of the broad H$\alpha$ emission line approximately 30 times higher than the values observed in LRD spectra.

Here, we emphasize that our approach does not rule out the existence of host galaxies in LRDs, 
but demonstrates that the SED model focusing only on AGNs successfully captures the characteristic spectral properties of LRDs:
(1) the v-shaped SED in rest-frame UV to optical bands and (2) modest NIR emission.
In our AGN-only spectral model, as a natural consequence, the red optical continuum is attributed to the attenuated AGN light.
This interpretation is indeed consistent with observations that the observed continuum and ${\rm H}\alpha$ line fluxes 
for LRDs follow the empirical correlation observed in nearby AGNs \citep{Greene_Ho_2005}.
In other words, if stellar light dominates the observed continuum flux of LRDs, the EW of broad ${\rm H}\alpha$ emission line 
would be unusually high, compared to normal AGNs.
This consistency is crucial, as it strengthens the interpretation of BH mass measurements using the single-epoch method.

\subsection{Variability (or not) of LRDs}

\citet{Kokubo_Harikane_2024} reported multi-epoch imaging analysis of five broad-line AGNs (including two unobscured and three LRD populations) over approximately $\sim 2-4$ epochs, finding no apparent variability features in any of these sources. This result poses a challenge to typical AGN scenarios, where a certain level of 
variability is expected. Several reasons for the lack of AGN variability have been proposed: under the AGN hypothesis, (i) an intrinsically non-variable AGN disk continuum, 
(ii) a host galaxy-dominated continuum, or (iii) scattering-dominated AGN emission, or alternatively, (iv) stellar activity producing broad Balmer emission lines.

We consider a scenario where apparent AGN variability is diminished due to multi-scale dust scattering, extending the above scenario (iii) to our model setup. 
In an extended dust envelope with a density gradient of $0<\gamma<1$ (as needed to explain the faint NIR emission in LRDs), the optical depth, or column density, profile becomes flattened, resulting in a dusty structure extending beyond $\sim 10$~pc to reach an extinction of $A_{V}=3$~mag.
In such an extended medium, photons emitted from the central source undergo scattering at a wide range of radius and scattered photons take various path lengths to 
reach out line of sight.
The delay time for scattered photons, given a typical scattering radius $r_{\rm scat}$, can be approximated as 
$\Delta t_{\rm scat}=r_{\rm scat}(1-\langle \cos \theta\rangle)/c\sim 10~{\rm yr}~(r_{\rm scat}/10~\pc)$, where $\langle \cos \theta\rangle$ is 
the scattering asymmetry factor (observed to be $\sim 0.6$ for forward scattering in the Orion Nebula; see \citealt{Saikia2018}).
This timescale is $\sim 30$ times longer than the typical AGN variability period, $t_{\rm var}\simeq 0.3~{\rm yr}$ \citep[][]{Burke2021}.
Assuming a spherical dust distribution and quasi-periodic AGN variability, this delay results in a damping factor of $t_{\rm var}/\Delta t_{\rm scat}\sim 1/30$
for the observed variability amplitude, due to phase cancelation \citep[][]{Daniel2017}. 
If the dust distribution is more centrally concentrated, as in standard dusty tori (i.e., $r_{\rm scat}\simeq r_{\rm sub}$ for $\gamma\gtrsim 2$), 
then $t_{\rm var}/\Delta t_{\rm scat}\sim 1/3$, leading to less significant damping of the observed variability amplitude.
As a result of this discussion, we expect that LRDs (or red quasars at lower redshifts) with bright NIR emission from hot dust tori
should show some detectable variability, if AGNs are intrinsically variable, without significant phase-canceling effects.
A more detailed quantitative analysis of this scenario will be explored in future work.

Recently, \cite{Zhang_2024} extended the variability analysis for a larger sample of 
$\sim 300$ LRDs obtained from publicly available photometric data.
These LRDs on average do not show detectable variability, but eight of them are identified
as variable LRD candidates with variability amplitudes of $0.24-0.82$ mag.
This non-zero fraction of variable LRDs ($\simeq 2.7\%$ of the parent sample) implies the potential contribution of super-Eddington accretion phases that would likely diminish AGN variability due to photon trapping in the disk (see Figure~15 in \citealt{Zhang_2024}).
The theoretical model for the variability feature are discussed in \cite{InayoshiX_2024}.

\subsection{Why do LRDs emerge at high redshifts of $z>4$\\ with a universal SED shape?}\label{sec:dustdistr}

There is a critical question: {\it Why do LRDs emerge at $z> 4$ with a uniform extinction level of $A_{\rm V}=3$ mag?} 
There is no clear answer to this question at this time, but some speculations are given below.

The key assumption in our model is the relaxation of density concentration around the AGN. 
This idea implies that LRDs form in the early stages of galaxy and AGN formation, characterized by dynamically evolving systems 
(e.g., clumpy gaseous structures, irregularly shaped and misaligned galactic disks, significant perturbations from mergers) 
before a well-organized dust torus structure is established \citep[e.g.,][]{Ceverino_2010,Ceverino_2015,Hopkins_2024}.
In the nascent stages of galaxy formation, the unsettled nature of the gas distribution permits a variety of attenuation 
levels and column densities. 
At higher redshifts, if the column density is typically higher and spans a broader range, numerous objects will exhibit $A_{\rm V}\gg 1$.
However, in the context of our proposed model, only those with moderate extinction levels with $A_{\rm V}\sim 1-3$ would reveal the characteristic SED shape,
as seen in observed LRD samples \citep[e.g.,][]{Barro_2023,Matthee_2024,Greene_2024, Kocevski_2024}.
With $A_{\rm V}\gtrsim 5$, which corresponds to $A_{3000}=1.6$ at $\lambda=3000~{\rm \AA}$
with the Orion-Nebula extinction curve, it is challenging to observe the rest-frame UV part of LRDs even with the JWST sensitivity. 
In this case, these objects appear to have only a red continuum without an UV excess, which are categorized as 
hot-dust-obscured AGNs observed at the lower-redshift universe \citep{AlonsoHerrero_2006,RamosAlmeida_2011,Ichikawa_2014,Hickox_2018,Lyu_2022}. We also note that these sources without a UV excess are quite rare at high redshift  \citep[][]{Barro_2023,Akins_2024}, approximately 95\% of the high-$z$ compact objects with red rest-frame optical continua have a blue UV excess \citep{Kocevski_2024}.
Our scenario seems more relevant in earlier epochs before a robust dust torus structure forms in its galactic nucleus.
When a centrally concentrated dust torus is established, as seen in nearby AGNs, there is a dichotomy between type 1 and type 2 AGN populations
due to the geometric arrangement 
\citep[e.g.,][]{RamosAlmeida_2017,Hickox_2018} rather than a modest variation due to $A_{\rm V}$.

Another critical element in our model is the flattening of dust extinction curves at shorter wavelengths, 
$\lambda \lesssim 3000~{\rm \AA}$.
This characteristic may help explain why dust-reddened objects with v-shaped SEDs predominantly appear at high redshifts.
In the early stages of protogalaxies, particularly those younger than one billion years ($z\gtrsim 6-7$), 
dust production is primarily driven by massive stars that expel heavy elements into the interstellar medium (ISM) 
through core-collapse supernovae \citep{Todini_Ferrara_2001,Valiante_2009}.
Due to effective dust destruction mechanisms that target smaller grains, the dust-size distribution tends to favor 
larger particles \citep{Nozawa_2006,Asano_2013}.
These larger grains, once released into the ISM, undergo shattering processes and break into smaller fragments in a typical 
timescale of $\gtrsim 0.5-1$ Gyr.
As a result, the extinction curve at shorter wavelengths begins to evolve towards those observed in the ISM of the Milky Way. 
The temporal evolution in dust characteristics would be critical for shaping the unique spectral features of LRDs 
only at high redshifts.

\subsection{Implications to the cosmological BH evolution}

In this paper, we propose that LRDs originate from rapidly growing BH embedded within 
dusty flows with an extended density distribution.
In this framework, we reach two intriguing conclusions regarding the cosmological BH evolution.

First, the luminosity of LRDs is primarily powered by AGNs that host accreting BHs.
These LRDs are more abundant than X-ray selected AGNs, suggesting an increase in the cosmic growth rate of BHs beyond $z\simeq 6$
\citep{Kokorev_2024,Kocevski_2024,Akins_2024}. 
The BH accretion rate density derived from the bolometric luminosity functions of LRDs is significantly higher 
than that estimated from other AGN surveys.
To reconcile this high value with the BH mass density at $z\sim 5$, a radiative efficiency of $\gtrsim 20\%$ -- twice the canonical
10\% value -- is required with a $3\sigma$ confidence level \citep{Inayoshi_Ichikawa_2024}.
This high radiative efficiency implies that the BHs are rapidly spinning in the LRD stage through violent mass accretion
during the early galaxy assembly.

Second, when the extinction level is accurately determined from the reddened optical continuum shape, 
the AGN continuum and broad-line luminosity can be more precisely estimated. 
These luminosity values are crucial for measuring BH mass using the single-epoch method.
For instance, \citet{Matthee_2024} calculated BH masses for LRDs using the ${\rm H}\alpha$ luminosity as an indicator 
of AGN continuum luminosity, but without accounting for dust reddening correction.
According to the empirical relationship in the single-epoch method, we derive that the Eddington ratio follows
$\lambda_{\rm Edd}\propto L_{\rm H\alpha}^{0.394}$ \citep[see][]{Greene_Ho_2005,Matthee_2024}.
Therefore, by applying a dust correction, the Eddington ratio increases by a factor of 
$\mathscr{L}\sim 10^{0.14A_{\rm V}}\sim 2.6$ for $A_{\rm V}=3$.
With this correction, the Eddington ratios of BHs reported by \citet{Matthee_2024} approach unity, 
suggesting that a significant fraction of LRDs may be super-Eddington accretors. 
This result highlights the importance of accurate dust correction for high-redshift AGNs,
particularly LRDs, to understand the true nature of their accretion dynamics (\citealt{Lupi_2024}; see also \citealt{Du_2014}).

\section{Conclusions}

JWST observations have uncovered the presence of very compact and red objects at the high-redshift universe,
referred to as little red dots (LRDs).
Spectroscopic observations have confirmed the broad components of emission lines and the presence of AGNs within LRDs.
Despite the breakthrough discovery, the characteristic v-shape spectral feature observed through JWST/NIRCam challenges us to 
understand the origin of these LRDs and the contribution from AGNs powered by accreting BHs.

In this work, we propose a SED model for LRDs spanning from rest-frame UV to infrared bands.
We hypothesize that the incident radiation from an AGN, characterized by a typical SED, is embedded in an extended dusty medium, 
which has an extinction law similar to those seen in dense regions such as Orion Nebula or certain AGN environments.
The UV-optical spectrum is described by dust-attenuated AGN emission with $A_{\rm V}\simeq 3$ mag, featuring a red optical continuum 
at $\lambda>4000~{\rm \AA}$ and a flat UV spectral shape established through a gray extinction curve at $\lambda<3000~{\rm \AA}$, 
due to the absence of small-size grains.
In the infrared, the SED is shaped by an extended dust and gas distribution ($\gamma<1$; $\rho\propto r^{-\gamma}$)
with a characteristic gas density of $\simeq 10-10^3~\cc$ at the dust sublimation radius, which allows relatively cool 
dust temperatures to dominate the radiation, thereby shifting the energy peak from near- to mid-infrared bands.
This model, unlike the typical AGN hot torus models, can produce a infrared SED flattening that is consistent with 
LRD observations through JWST MIRI.
In this scenario, there is no need for additional stellar emission or AGN scattered light.

\acknowledgments
We greatly thank Changhao Chen, Seiji Fujimoto, Jinyi Shangguan, and Mingyang Zhuang for constructive discussions. 
We acknowledge support from the National Natural Science Foundation of China (12073003, 12003003, 11721303, 11991052, 11950410493), 
and the China Manned Space Project (CMS-CSST-2021-A04 and CMS-CSST-2021-A06). 
This work is also supported by Japan Society for the Promotion of Science (JSPS) KAKENHI (20H01939; K.~Ichikawa). LCH was supported by the National Science Foundation of China (11991052, 12233001), the National Key R\&D Program of China (2022YFF0503401), and the China Manned Space Project (CMS-CSST-2021-A04, CMS-CSST-2021-A06).
This work made use of Astropy:\footnote{http://www.astropy.org} a community-developed core Python package and an ecosystem of tools and resources for astronomy \citep{astropy:2022}.
This work use the dust extinction data from version 23.01 of \texttt{CLOUDY}, last described by \cite{CLOUDY_23}.

\bibliography{ref}{}
\bibliographystyle{aasjournal}

\end{document}